\documentclass[a4paper,10pt]{article}

\usepackage[latin1]{inputenc}

\usepackage[fleqn]{amsmath} 

\usepackage{amsfonts}
\usepackage{amssymb}
\usepackage[english]{babel}
\usepackage{comment}
\usepackage{graphicx}
\usepackage{rotating}
\usepackage{booktabs}

{\left\lbrace\begin{array}{@{}l@{}}}%
{\end{array}\right.}

\DeclareMathOperator{\DFT}{DFT}
\DeclareMathOperator{\RDFT}{RDFT}
\DeclareMathOperator{\CDFT}{CDFT}
\DeclareMathOperator{\DCT}{DCT}
\DeclareMathOperator{\DST}{DST}

\DeclareMathOperator{\flop}{flop}
\DeclareMathOperator{\mul}{mul}
\DeclareMathOperator{\add}{add}

\begin{document}

\title{Improved QFT algorithm for power-of-two FFT}


\author{Lorenzo Pasquini\footnote{\em Falconara Marittima (AN), Italy, pasquini.paper@gmail.com}} 
 


\maketitle

\begin{abstract} 
This paper shows that it is possible to improve the computational cost, the memory requirements and the accuracy of Quick Fourier Transform (QFT) algorithm for power-of-two FFT (Fast Fourier Transform) just introducing a slight modification in this algorithm.
The new algorithm requires the same number of additions and multiplications of split-radix 3add/3mul, one of the most appreciated FFT algorithms appeared in the literature, but employing only half of the trigonometric constants. 
These results can elevate the QFT approach to the level of most used FFT procedures. 
A new quite general way to describe FFT algorithms, based on signal types and on a particular notation, is also proposed and used, highligting its advantages.

\end{abstract} 




\section{Introduction} \label{sec:introduction}

The Fast Fourier Transform (FFT) is a basic subject in signal processing, and many FFT algorithms have been proposed in literature \cite{Duhamel_Vetterli_1990} to compute it.
An ideal FFT algorithm should have many desidered characteristics, according to the applicative context (the most important usually are: low computational cost, low memory requirements, high numerical accuracy, simplicity) but, up to date, no FFT algorithm is optimal in all these characteristics.
For this reason new FFT algorithms with a different compromise between these desired characteristics are welcome also if they haven't got the best theoretical computational cost.
For lenght $N=2^r$ the most popular algorithm is radix-2 \cite{Cooley_Tuckey_1965}, while a very appreciated algorithm is split-radix \cite{Duhamel_Hollmann_1984}, \cite{Martens_1984}, \cite{Vetterli_Nussbaumer_1984}, of whom some interesting variants exist \cite{Bouguezel_2007}, \cite{Kamar_Elcherif_1989}.
However some algorithms more efficient than split-radix, if the used computational model evaluates efficiency with required flops (floating point operations), recently appeared: the scaled split-radix \cite{Johnson_Frigo_2007} (also called tangent FFT \cite{Bernstein_2007}), the scaled odd-tail \cite{Lundy_Van_Buskirk_2007}, and other ones are possible \cite{haynal:generating}.
Another good algorithm is the Quick Fourier Transform \cite{Guo_Sitton_qft_1994} (called `classical QFT algorithm' in this paper), a real factor algorithm that uses few trigonometric constants, and has a good (not excellent) computational cost.
In this paper we show how to improve the computational cost, the memory requirements (using the same few trigonometric constants), and numerical accuracy of classical QFT for $N=2^r$, modifying this algorithm with the addition of a further but appropriate intermediate decomposition. 
Characteristics of the new algorithm, called improved QFT, make it a good choise for fixed point implementation, where it is a good alternative to split-radix 3mul-3add.
In order to point out the reason of this improvements, we introduce a new way, based on signal types and on a particular notation, to describe FFT algorithms.
This new approach has many other advantages too (highligted in appendix), inherent many steps of `algorithms life': the research and the theorical developing of algorithms, the exposition of algorithms, and the implementation of algorithms. 
Let us briefly summarize the content of the paper. 
In sect.2 we show the new way to decribe FFT algorithms that we use in this paper. 
In sect.3 the reader can find the used basic elaborations. 
In sect.4 and sect.5 we describe both classical and improved QFT algorithm respectively.
Finally, in sect.6 we discuss the memory requirement, the computational cost ad the accuracy relative to the proposed improved QFT algorithm.


\section{a new approach to describe FFT algorithms}

A new manner to describe FFT algorithms is used in this paper, and we could use it for many other power-of-two FFT algorithms (quite all the ones that use the `divide et impera' approach).
This new approach is made of these components:
\begin{itemize}

\item 
use of new concepts: signal types, non-zero value time indeces $nze\_n$, stored $sto\_n$ indeces, stored $sto\_k$ indeces, independent-value stored harmonics $ind\_sto\_k$, storage size $ln$ and $lk$ parameters of a signal in temporal and frequency domain respectively.
Some of these concepts are useful since FFT algorithms described in this paper create many descendent signals of whom we have to compute a pruned input and/or output transform.

\item
use of a mnemonic notation to describe relevant characteristics of signals created inside the FFT algorithm.

\item
use of a table (as Tab.\ref{tab:notation}) to describe the characteristics of any signal type used in an algorithm. 
We should look at this table while we read this paper.

\item
use of a table (as Tab.\ref{tab:implementation}) that describes the matching between each signal and the array cells that store the signal, in an implementation of the algorithm in a suitable programming language (useful if we want to write the code of this FFT algorithm).

\item
use of the decomposition tree (as Fig.\ref{fig:classical_qft}), a graphical representation that shows both the concatenation of basic elaborations used by the functions, and the signal types handled by an FFT algorithm.

\end{itemize}


\subsection{Basic definitions} \label{sec:definitions}

\begin{itemize}



\item 
\emph{Signal (time) periodization $N$}:
the time period of the fundamental frequency of the transform applied to a signal.

 

\item
\emph{$\DFT$, $\DCT$, $\DST$ transforms:}
three different ways to represent a signal by superposing stationary-amplitude and stationary-frequency oscillations. 
Analytically speaking we can define them as follows:
\begin{equation} \label{equ:qft_0A}
S(k)=\DFT[s](k)= \sum_{n=0}^{N-1} s(n) \cdot e^{-i \theta \cdot n \cdot k} \quad k \in \{0,1,2,\dots,(N-1)\} 
\end{equation}
\begin{equation}  \label{equ:qft_0B}
S(k)=\DCT[s](k)= \sum_{n=0}^{\frac{N}{2}} s(n) \cdot \cos( \theta \cdot n \cdot k) \quad k \in \{0,1,2,...,(\frac{N}{2})\}
\end{equation}
\begin{equation}  \label{equ:qft_0C}
S(k)=\DST[s](k)= \sum_{n=1}^{\frac{N}{2}-1} s(n) \cdot \sin( \theta \cdot n \cdot k) \quad  k \in \{1,2,3,...,(\frac{N}{2}-1)\}
\end{equation}
where $\theta$ is the angle pulse of fundamental frequency and defined as:
\begin{equation} \label{equ:qft_0D}
\theta=  \frac{2 \cdot \pi}{N}
\end{equation}
and $N$ is the periodization of the transform we apply to the signal.
The $\DCT$ and $\DST$ transforms are defined in compliance with the definitions given in \cite{Guo_Sitton_qft_1994}, with the only difference that here we describe them in terms of the periodization $N$  (while $N$ is the half-periodization in \cite{Guo_Sitton_qft_1994}).
We can call them $\DCT-0$ and $\DST-0$, to distinguish them from other $\DCT$ and $\DST$ types already defined in literature (however they are similar to $\DCT-I$ and $\DST-I$ respectively).
Let us observe that for $\DFT$ the concept of periodization coincides with usual lenght term.
Moreover we can apply $\DFT$ transform both to real ($\RDFT$) and complex ($\CDFT$) valued signals.
With an abuse of notation, we will use the $\DFT$, $\DCT$, $\DST$ terms in case of pruned input and/or output too (when only a subset of $N$ values $s(n)$ are non-zero, or when only a subset of $N$ values $S(k)$ are required).


\item
\emph{conversion}: the elaboration resulting in the attainment of an only child signal from the mother one.
An ideal conversion used inside a FFT algorithm doesn't increase the number of indeces $n$ or $k$ to handle, passing from mother to child signal, and requires a few flops.

\item
\emph{Decomposition}:
the elaboration resulting in the attainment of two or more children signals from the mother one.
An ideal decomposition used inside a FFT algorithm doesn't increase the total number of indeces $n$ or $k$ to handle, passing from mother to children signals (thus each child signal has both less $n$ and less $k$ indeces to handle versus its mother), and requires a few flops.

\item
\emph{Forward phase of a conversion or of a decomposition}:
this is the phase where the time elements are processed. Therefore, when we apply it, the known elements are those of the mother signal, and the unknown elements those of the child signal (or signals).

\item
\emph{Backward phase of a conversion or of a decomposition}:
this is the phase where the frequency-domain elements are processed. Therefore, when we apply it, the known elements are those of the child signal (or signals), and the unknown elements those of the mother signal.

\item
\emph{non-zero value time indices grouping $nze\_n(s)$ of a signal $s$}:
the group of $n$ indeces where $s(n)$ has not a-priori known zero value ($nze=$ non-zero).
This definition is useful since FFT algorithms described in this paper create many descendent signals that contain many a-priori known temporal zero-valued samples (pruned input) and thus of whom we compute a pruned transform.
It is therefore evident that, in correspondence of related time instants, it is not required to apply any processing or to consume memory, improving the overall algorithm efficiency. 
For these reasons a processing that handles only residual time indeces can be used.

\item
\emph{stored $sto\_n(s)$ grouping of a signal $s$}:
the grouping of only indices $n$ whose $s(n)$ value we need (or it is convenient) to store in memory (and thus to compute too, if they are unknown).
In general case, $sto\_n$ and $nze\_n$ groupings can differ if a signal contains some dependent (as identical, or opposite) values $s(n)$, as it happens when we handle symmetric or antisimmetric signals (in fact, in this case, we can store only independent values, that means only a subset of $nze\_n$).
However, the $sto\_n$ grouping coincides both with $nze\_n$ grouping and with the group of indipendent temporal elements, in any signal used in this paper.

\item
\emph{stored $sto\_k(s)$ grouping of a signal $s$}:
the grouping of only indices $k$ whose frequency-domain components we need (or it is convenient) to store in memory (and thus to compute too inside the FFT algorithm).
In particular, for any $k$ external the $sto\_k$ grouping, one of this two conditions happens:
\begin{itemize}

\item 
we are not interested in calculating the corresponding frequency-domain component (that can be different from zero), for example since it can be obtained from $S(k)$ values of stored harmonics, without any further computation.

\item
we are interested in calculating the corresponding frequency-domain component, but we obtain its value from the values of frequency-domain components in other stored harmonics of the same signal, without reserving array cells to store them.
For example, if $\RDFT[s]$ computation is the goal, than $k=N-1$ is not a required harmonic, since we can obtain its complex frequency-domain component $S(k=N-1)$ from $S(k=1)$
(in this case, storing the second half of $\RDFT$ frequency-domain signal too, is only for completeness in exposition of result, not a necessity of FFT algorithm).

\end{itemize}
This definition is useful since FFT algorithms described in this paper create many pruned-output descendent signals.

\item
\emph{indipendent-value stored harmonics grouping $ind\_sto\_k(s)$ of a signal $s$}:
the subset of $sto\_k(s)$ group (created selecting $k$ indeces starting from $k=min(sto\_k)$ and then increasing $k$) where any associated $S(k)$ value has at least a real component that is independent from the other ones of this group.
This grouping coincides with $sto\_k$ grouping for any signal type used in this paper, except for three signal types, used in cassical QFT, as we will see in sect. \ref{sec:notation_sub}.
This concept is needed to catch the reasons of inefficiences of classical QFT versus improved QFT.

\item
\emph{storage size $ln(s)$ of a signal $s$ in temporal domain}: 
the number of array cells, with real values, used to store the temporal $s(n)$ values, with $n \in sto\_n(s)$, of handled signal $s$.
For any signal used in this paper $ln$ is univocally determined by the $sto\_n(s)$ group, according to these relations:
\begin{equation} \label{equ:qft_1A}
ln(s) = \left\{ \begin{array}{ll}
card(sto\_n(s)) \quad &  \text{for $\RDFT$, $\DCT$, $\DST$} \\
2 \cdot card(sto\_n(s)) \quad & \text{for $\CDFT$}  \\
\end{array} \right.
\end{equation}
where $card(A)$ is the cardinality of the set $A$.

\item
\emph{storage size $lk(s)$ of a signal $s$ in frequency domain}: 
the number of array cells, with real values, used to store the frequency-domain $S(k)$ values, with $k \in sto\_k(s)$ of handled signal $s$.
For any signal used in this paper $lk$ is univocally determined by the $sto\_k(s)$ group, according to these relations:
\begin{equation} \label{equ:qft_1B}
lk(s) = \left\{ \begin{array}{ll}
card(sto\_k(s)) \quad &  \text{for $\DCT$, $\DST$} \\
2 \cdot card(sto\_k(s)) \quad & \text{for $\CDFT$}  \\
2 \cdot [card(sto\_k(s))-1] \quad & \text{for $\RDFT$}  \\
\end{array} \right.
\end{equation}

\item
\emph{signal type}: the configuration of characteristics of a signal that contains any information we need to know about a signal both to choise the theoretical basic elaboration to apply to it, and to write the software code we apply to this signal inside the FFT algorithm (thus each signal type is handled always with the same basic elaboration, in a FFT algorithm).
In this paper the applied transform, $sto\_n$, $sto\_k$ groupings are the only informations required to determine a signal type.
Differently, the periodization $N$ of a signal doesn't contribute to determinate the `signal type' (two signals can belong to the same `signal type', even if they have different periodization $N$), since the recursive function we apply to a signal doesn't depend on periodization $N$.
Tab.\ref{tab:notation} lists any signal type used in this paper.

\item
\emph{Decomposition Tree}: a graphical representation of the FFT algorithm.
This representation has the advantage that lets us to know both the structure of the algorithm and any relevant characteristic of any descendent signal created by the concatenation of basic elaborations used to build the whole FFT algorithm, hiding mathematical details of these basic elaborations (that we can see in a separate section). 
The usage of word `tree' is justified by the fact that quite all power-of-two FFT algorithms, including the innovative one that will be described in the following, basically convert and decompose the original (root) signal into different other descendent signals, thus giving origin to a decomposition tree (see for instance  Figs. \ref{fig:classical_qft},\ref{fig:improved_qft}).

\end{itemize}


\subsection{Signal types and notation} \label{sec:notation}

\subsubsection{Signal types}

In this paper we show that applying a small modification to the classical QFT, some of its characteristics improve.
Catching the reasons of this fact is not simple, and requires to analyze the details of the computations, and of created signals, used step by step in the classical and in the new algorithm. 
In particular, we need to focus on $sto\_n$ and $sto\_k$ groups of some gradually created signals.
In fact, as we shall see, in the two algorithms shown in this paper, sometimes we use signals with identical characteristics. 
Other times we create some signals to whom we apply the same transform (i.e. $\DCT$) and whose $sto\_n$ and $sto\_k$ groupings differ only by the presence or not of an only $sto\_n$ index, or of an only $sto\_k$ harmonic.
We need to focus on these slight differences since they are relevant to make the new algorithm more efficient than the classical one.
For this reason a new manner to describe FFT algorithms, that focus on the `signal type' concept is helpful.
It particular, (as already announced in sect. \ref{sec:definitions}) limitating the analysis to the two algorithms we are going to describe, we need to compare (and to focus on) three signal characteristics at the same: $sto\_n$,  $sto\_k$, and the applied transform, to determine the `signal type' of a signal.
The idea of formalize the concept of signal type, and to associate an unique signal name to it in a sistematic way, in description of the algorithms, shows many other advantages in development, exposition and implementation of algorithms. 
Details of these advantages are shown in appendix.
Let us stress that signal types used inside an algorithm are not a-priori known, and can be determined only analyzing the mathematical details of any used basic elaborations, as we will see in sect. \ref{sec:conversions}.

\subsubsection{Notation for signal types } \label{sec:notation_sub}

Instead of associating a casual name to each signal type, we prefer to use a mnemonic descriptive notation, that identifies each signal type by means of a suitable sequence of symbols (see Tab.\ref{tab:notation}), according to these rules:
\begin{itemize}

\item
the main symbol is `s' (s=signal) in any case.

\item
the first subscript symbol identifies the applied transform: `$cx$'  (complex), `$re$' (real), `$dc$' ($\DCT$), `$ds$' ($\DST$).

\item
the second subscript symbol refers to $sto\_n$: `$o$' means generic odd, `$e$' or `$e_1$' are two different grouping of only even $n$ indices,  `$t$' or `$t_1$' (generic total) are two different grouping of both even and odd $n$ indices.

\item
the third subscript symbol refers to $sto\_k$: `$o$' (generic odd), `$e$' (generic even), `$t$' or `$t_1$' (generic total).

\end{itemize}
This notation highlights the parallelism in the elaboration used in the corresponding recursive functions, in the $\DCT$  context, and in the $\DST$ context, inside classical, as improved, QFT algorithms. 
In this way, for many functions (except in the $dct\_to\_cla$ or $dst\_to\_cla$ functions used in the classical QFT), we can switch between signal types used in $\DCT$ context, to the ones used in $\DST$ context of the same algorithm, simply replacing the `$dc$' by the `$ds$' subscript.
As a side effect of this notation, there is no bijective correspondence among a single subscript symbol, and a single feature of the signal (except fot the 1st subscript), but only among a sequence of subscript symbols, and a signal type.
For example, the subscript `$e$' referring to $sto\_k$ identifies:
\begin{itemize}

\item
the group $sto\_k=\{0,2,\dots,(\frac{N}{2})\}$ if it is used in $s_{dc\_te}$ sequence of symbols.

\item
the group $sto\_k=\{2,4, \dots, (\frac{N}{2}-2)\}$ if it is used in $s_{ds\_te}$ sequence of symbols.

\end{itemize}
Notice that the exposed notation for signal types does not require to distinguish the `$t$' symbol (or any other symbol) depending on whether it refers to the grouping $sto\_n$, or it refers to the grouping  $sto\_k$ (for example using the $t_n$ in the first case, and the symbol $t_k$ in the second case), because we only need to consider the position of the symbol in the notation to see if it relates to $sto\_n$ or $sto\_k$. 
This choice has the advantage to make the name of each signal shorter.
Moreover this notation has the advantage that reading a signal type name we can immediately remember many characteristics of this signal.
For instance, reading the term $s_{ds\_t_{1} o}$, we remember that it denotes the signal type to whom we apply the $\DST$, having both some even and odd residual time $n$ indices, and for which only some odd $k$ harmonics are required.
Tab.\ref{tab:notation} reports all and only the sequences of symbols (signal types) used in this paper (it must be also noted that only some of the feasible symbol combinations describe signal types effectively occurring in the addressed algorithms). 
Let us observe that $s_{dc\_t1e}$, $s_{dc\_t1t}$, $s_{ds\_t1o}$ are the only used signal types with $lk=ln+1$.
It means they store a real $S(k)$ element dependent from the other ones of the same $sto\_k$ group (and thus that $ind\_sto\_k \neq sto\_k$ holds), because a pruned input signal with only $ln$ real independent temporal elements, can have a maximum of $ln$ independent real transformed elements.

These three signal types are used only in classical QFT, and some of them cause inefficiencies, as we will see in sect. \ref{sec:memory_old_qft} and \ref{sec:memory_new_qft}.

\subsubsection{notation for signals} \label{sec:notation_sub_2}

Each used signal is described by means of a notation that slightly modifies the notation used for the associated signal type, according to these rules:
\begin{itemize}
\item 
the 1st symbol is `$s$' for temporal signals, and `$S$' for frequency-domain signals.

\item
an optional subscript identifier (numbers and/or capital letters), can be inserted after the 1st `$s|S$' symbol, to distinguish the handled signal from other signals of the same type used in the same context.

\end{itemize}
For example $s_{dc\_tt}$ and $s_{A\_dc\_tt}$, $s_{3,1\_A\_dc\_tt}$ are three different temporal signals of the same type `$s_{dc\_tt}$', while $S_{ds\_ot}$ and $S_{A\_ds\_ot}$, $S_{4,7\_A\_ds\_ot}$ are three different frequency-domain signals of the same type `$s_{ds\_ot}$'.

\begin{table}[tb]
\caption{Transform type, $sto\_n$, $sto\_k$ groups, storage size $ln$ and $lk$ parameters of a signal in temporal and frequency domain respectively, associated to any signal type used in this paper.}
\label{tab:notation}
\centering
\scalebox{0.8}
{
\begin{tabular}{cccccc}
\toprule
signal type & transform type & sto\_n=nze\_n & sto\_k & ln & lk\\
\midrule
$s_{cx\_tt}$ & $\CDFT$ & $\{ 0,1,2,\dots,(N-1) \}$ & $ \{ 0,1,2,\dots,(N-1) \}$ & $2 \cdot N$ & $2 \cdot N$ \\
$s_{re\_tt}$ & $\RDFT$ & $\{ 0,1,2,\dots,(N-1) \}$ & $ \{ 0,1,2,\dots,(\frac{N}{2}) \}$  & $N$ & $N$ \\
$s_{dc\_tt}$ & $\DCT$ & $\{ 0,1,2,\dots,(\frac{N}{2}) \}$ & $ \{ 0,1,2,\dots,(\frac{N}{2}) \}$  & $N/2+1$ & $N/2+1$ \\
$s_{dc\_et}$ & $\DCT$ & $\{ 0,2,4,\dots,(\frac{N}{2}) \}$ & $\{ 0,1,2,\dots,(\frac{N}{4}) \}$  & $N/4+1$ & $N/4+1$ \\
$s_{dc\_ot}$ & $\DCT$ & $\{ 1,3,5,\dots,(\frac{N}{2}-1) \}$ & $ \{ 0,1,2,\dots,(\frac{N}{4}-1) \}$  & $N/4$ & $N/4$ \\
$s_{dc\_te}$ & $\DCT$ & $\{ 0,1,2,\dots,(\frac{N}{4}) \}$ & $ \{ 0,2,4,\dots,(\frac{N}{2}) \}$  & $N/4+1$ & $N/4+1$ \\
$s_{dc\_to}$ & $\DCT$ & $\{ 0,1,2,\dots,(\frac{N}{4}-1) \}$ & $ \{ 1,3,5,\dots,(\frac{N}{2}-1) \}$  & $N/4$ & $N/4$ \\
$s_{dc\_oe}$ & $\DCT$ & $ \{ 1,3,5,\dots,(\frac{N}{4}-1) \}$ & $ \{ 0,2,4,\dots,(\frac{N}{4}-2) \}$ & $N/8$ & $N/8$\\
$s_{dc\_oo}$ & $\DCT$ & $\{ 1,3,5,\dots,(\frac{N}{4}-1) \}$ & $\{ 1,3,5,\dots,(\frac{N}{4}-1) \}$ & $N/8$ & $N/8$\\
$s_{dc\_t_{1} e}$ & $\DCT$ & $\{ 0,1,2,\dots,(\frac{N}{4}-1) \}$ & $\{ 0,2,4,\dots,(\frac{N}{2}) \}$ & $N/4$ & $N/4+1$\\
$s_{dc\_t_{1} t}$ & $\DCT$ & $\{ 0,1,2,\dots,(\frac{N}{2}-1) \}$ & $\{ 0,1,2,\dots,(\frac{N}{2}) \}$ & $N/2$ & $N/2+1$\\
$s_{ds\_tt}$ & $\DST$ & $\{ 1,2,3,\dots,(\frac{N}{2}-1) \}$ & $ \{ 1,2,3,\dots,(\frac{N}{2}-1) \}$ & $N/2-1$ & $N/2-1$\\
$s_{ds\_et}$ & $\DST$ & $ \{ 2,4,6,\dots,(\frac{N}{2}-2) \}$ & $ \{ 1,2,3,\dots,(\frac{N}{4}-1) \}$ & $N/4-1$ & $N/4-1$\\
$s_{ds\_te}$ & $\DST$ & $\{ 1,2,3,\dots,(\frac{N}{4}-1) \}$ & $ \{ 2,4,6,\dots,(\frac{N}{2}-1) \}$ & $N/4-1$ & $N/4-1$\\
$s_{ds\_to}$ & $\DST$ & $ \{ 1,2,3,\dots,(\frac{N}{4}) \}$ & $ \{ 1,3,5,\dots,(\frac{N}{2}-1) \}$ & $N/4$ & $N/4$\\
$s_{ds\_ot}$ & $\DST$ & $\{ 1,3,5,\dots,(\frac{N}{2}-1) \}$ & $\{ 1,2,3,\dots,(\frac{N}{4}) \}$ & $N/4$ & $N/4$\\
$s_{ds\_oe}$ & $\DST$ & $\{ 1,3,5,\dots,(\frac{N}{4}-1) \}$ & $ \{ 2,4,6,\dots,(\frac{N}{4}) \}$ & $N/8$ & $N/8$\\
$s_{ds\_oo}$ & $\DST$ &  $\{1,3,5,\dots,(\frac{N}{4}-1)\}$ & $ \{ 1,3,5,\dots,(\frac{N}{4}-1) \}$ & $N/8$ & $N/8$\\
$s_{ds\_t_{1} o}$ & $\DST$  & $\{ 1,2,3,\dots,(\frac{N}{4}-1) \}$ & $ \{ 1,3,5,\dots,(\frac{N}{2}-1) \}$ & $N/4-1$ & $N/4$\\
$s_{ds\_e_{1} o}$ & $\DST$  & $(\frac{N}{4})$ & $1$ & $1$ & $1$\\
\bottomrule
\end{tabular}
}
\end{table}


\section{some basic elaborations (decompositions and conversions) shared by classical and improved QFT} \label{sec:conversions}

Algorithms developed in this paper involve some common decompositions and conversions, but applied to different signal types: separation of even harmonics from odd ones, separation of even time indices from odd ones, even harmonics halving, even time indices halving. 
It must be noted that, in the time-domain case study (forward phase) such decompositions  or conversions take the time samples of mother-signal as known data, whereas those of derived signals as unknown. 
On the contrary, in the frequency-domain case study (backward phase) the relationship between known-unknown data and samples of mother-derived signals is inverted.
Moreover we describe each elaboration not referring to a specific signal type, since each basic elaboration is applied to many different signal types in this paper.


\subsection{Separation between even and odd time indices}  \label{sec:separe_n}

Let $s_{t_n}$ be the generic mother signal and $s_{o_n}$, $s_{e_n}$ the two created children signals.
The temporal analytical equations corresponding to the separation between even and odd time indices are targeted to separate the only $nze\_n=sto\_n$ indices of mother signal types (to which this decomposition is applied):
\begin{equation} \label{equ:qft_5A}
s_{e_n}(n) = \left\{ \begin{array}{ll}
s_{t_n}(n) \quad &  \text{even} \, n, \, n \in nze\_n(s_{t_n}) \\
0 \quad & \text{otherwise}  \\
\end{array} \right.
\end{equation}
\begin{equation} \label{equ:qft_5B}
s_{o_n}(n) = \left\{ \begin{array}{ll}
s_{t_n}(n) \quad &  \text{odd} \, n, \, n \in nze\_n(s_{t_n}) \\
0 \quad & \text{otherwise}  \\
\end{array} \right.
\end{equation}
Within the $\DCT$ it can be easily proved that this decomposition generates the following equations (backward phase):
\begin{equation}
\begin{split} \label{equ:qft_5C}
 \DCT[s_{t_n}](k) &= \DCT[s_{e_n}](k) + \DCT[s_{o_n}](k) \\
& k \in sto\_k(s_{t_n}), \quad  k \in [0, \frac{N}{4})  \\
\end{split} \\
\end{equation}
\begin{equation}
\begin{split} \label{equ:qft_5D}
 \DCT[s_{t_n}](\frac{N}{2}-k) &= \DCT[s_{e_n}](k) - \DCT[s_{o_n}](k) \\
& k \in sto\_k(s_{t_n}), k \in [0, \frac{N}{4})  \\
\end{split} \\
\end{equation}
\begin{equation} \label{equ:qft_6} 
\DCT[s_{t_n}](k) = \DCT[s_{e_n}](k)  \quad k \in sto\_k(s_{t_n})  \cap \{\frac{N}{4}\} 
\end{equation}
Similarly, within the $\DST$ case, the same eq.(\ref{equ:qft_5C}),(\ref{equ:qft_5D}),(\ref{equ:qft_6}) hold swapping $s_{e_n}$ and $s_{o_n}$.
Here we list the children signal types (described in Tab.\ref{tab:notation}) obtained applying this decomposition to different mother signal types.

If $s_{t_n}=s_{dc \_tt}$ holds then $s_{o_n}=s_{dc \_ot}$ and $s_{e_n}=s_{dc \_et}$.

If $s_{t_n}=s_{ds \_tt}$ holds then $s_{o_n}=s_{ds \_ot}$ and e $s_{e_n}=s_{ds \_et}$.

If $s_{t_n}=s_{dc \_t_{1}t}$ holds then $s_{o_n}=s_{dc \_ot}$ and $s_{e_n}=s_{dc \_et}$.

Now we describe in details how to obtain the signal type of children signals, created by this basic elaboration, if the mother signal is $s_{t_n}=s_{dc\_tt}$, that has $sto\_n(s_{dc\_tt})=nze\_n(s_{dc\_tt})=\{0,1,2,\dots,\frac{N}{2}\}$, according to Tab.\ref{tab:notation}.
Using eq.(\ref{equ:qft_5A}),(\ref{equ:qft_5B}) we obtain $nze\_n(s_{e_n})=\{0,2,\dots,\frac{N}{2}\}$ and $nze\_n(s_{o_n})=\{1,3,\dots,\frac{N}{2}-1\}$.
Morever the mother signal has $sto\_k(s_{dc\_tt})=\{0,1,2,\dots,\frac{N}{2}\}$ according to Tab.\ref{tab:notation}. 
Thus eq.(\ref{equ:qft_5C}),(\ref{equ:qft_5D}),(\ref{equ:qft_6}) force us to know (and to store) $sto\_k(s_{e_n})=\{0,1,\dots,\frac{N}{4}\}$ and $sto\_k(s_{o_n})=\{0,1,\dots,(\frac{N}{4}-1)\}$, computed by means of $\DCT$ in both cases.
Combining these informations we obtain $s_{o_n}=s_{dc\_ot}$ and $s_{e_n}=s_{dc\_et}$, according to Tab.\ref{tab:notation}.
In a similar manner we can obtain the children signal types handled in the other cases of this basic elaboration, or in the other basic elaborations.


\subsection{Separation between even and odd harmonics} \label{sec:separe_k}

This elaboration is dual to the one described in sect.\ref{sec:separe_n}.
Let $s_{t_k}$ be the generic mother signal and $s_{e_k}$, $s_{o_k}$ the two created children output signals.
Within the $\DCT$ context, it can be easily proved that separation between even and odd harmonics generates the following time-domain relations:
\begin{equation} \label{equ:qft_1}
s_{e_k}(n) = \left\{ \begin{array}{ll}
s_{t_k}(n) + s_{t_k}(\frac{N}{2}-n) \quad & n \in nze\_n(s_{t_k}), \, n \in [0,  \frac{N}{4}-1]  \\
s_{t_k}(n) \quad &  \{n=\frac{N}{4}\}  \cap nze\_n(s_{t_k}) \\
0 \quad & \text{otherwise}  \\
\end{array} \right.
\end{equation}
\begin{equation} \label{equ:qft_2}
s_{o_k}(n) = \left\{ \begin{array}{ll}
s_{t_k}(n) - s_{t_k}(\frac{N}{2}-n) \quad & n \in nze\_n(s_{t_k}), \, n \in [0,  \frac{N}{4}-1]  \\
0    \quad & \text{otherwise}  \\
\end{array} \right.
\end{equation}
while, in the $\DST$ case, we obtain the same eq.(\ref{equ:qft_1}),(\ref{equ:qft_2}) swapping betweeen $s_{e_k}$ and $s_{o_k}$.
The frequency-domain analytical equations corresponding to such a decomposition  (backward phase) are
targeted to align the $sto\_k$ indices of children signals $s_{e_k}$ and $s_{o_k}$:
\begin{equation} \label{equ:qft_4B}
S_{t_k}(n) = \left\{ \begin{array}{ll}
S_{e_k}(k) \quad & \text{even} \, k, \, k \in sto\_k(s_{t_k}) \\
S_{o_k}(k) \quad & \text{odd} \, k, \, k \in sto\_k(s_{t_k}) \\
\end{array} \right.
\end{equation}
Here we list the children signal types (described in Tab.\ref{tab:notation}) obtained applying this decomposition to different mother signal types.

If $s_{t_k}=s_{dc\_tt}$ holds then $s_{o_k}=s_{dc \_to}$ and $s_{e_k}=s_{dc \_te}$.

If $s_{t_k}=s_{dc\_ot}$ holds then $s_{o_k}=s_{dc \_oo}$ and $s_{e_k}=s_{dc \_oe}$.

If $s_{t_k}=s_{ds\_tt}$ holds then $s_{o_k}=s_{ds \_to}$ and $s_{e_k}=s_{ds \_te}$.

If $s_{t_k}=s_{ds\_ot}$ holds then $s_{o_k}=s_{ds \_oo}$ and $s_{e_k}=s_{ds \_oe}$.



\subsection{Even Harmonics Halving} \label{sec:k_even} 

The generic mother signal $s_{e_k}$, characterized by periodization $N$, is converted into the child signal $s_{t_k}$, with time periodization $N_A=(\frac{N}{2})$, whose harmonics we are interested to are both even and odd, and are obtained by halving each even harmonic of  $s_{e_k}$, keeping unchanged their associated frequency-domain components.
It can be easily proved that this corresponds to the following temporal relation:
\begin{equation} \label{equ:qft_9}
s_{t_k}(n) = \left\{ \begin{array}{ll}
s_{e_k}(n) \quad &  n \in nze\_n(s_{e_k}) \\
0 \quad & \text{otherwise}  \\
\end{array} \right.
\end{equation}
From a frequency-domain perspective (backward phase), dependening on which transform we are interested to, in $\DCT$ context, the following relation holds:
\begin{equation} \label{equ:qft_10}
\DCT[s_{e_k}](k=2 \cdot k_A) = \DCT[s_{t_k}](k_A) \quad k \in sto\_k(s_{e_k})
\end{equation}
while in $\DST$ context the same eq.(\ref{equ:qft_10}) holds, changing $\DCT$ with $\DST$. 
Here we list the children signal types (described in Tab.\ref{tab:notation}) obtained applying this decomposition to different mother signal types.

If $s_{e_k}=s_{dc\_te}$ holds then $s_{t_k}=s_{dc\_tt}$.

If $s_{e_k}=s_{dc\_oe}$ holds then $s_{t_k}=s_{dc\_ot}$.

If $s_{e_k}=s_{ds\_te}$ holds then $s_{t_k}=s_{ds\_tt}$.

If $s_{e_k}=s_{ds\_oe}$ holds then $s_{t_k}=s_{ds\_ot}$.

If $s_{e_k}=s_{dc\_t_{1}e}$ holds then $s_{t_k}=s_{dc\_t_{1}t}$.


\subsection{Even Time Indices Halving} \label{sec:n_even} 

This elaboration is dual to the one described in sect. \ref{sec:k_even}.
The generic mother signal $s_{e_n}$, containing only some even time indices, is converted into the child signal $s_{t_n}$, with periodization $N_A=(\frac{N}{2})$, with both even and odd time indices.
The signal $s_{t_n}$ is obtained by halving any even $n$ index of $s_{e_n}$, keeping unchanged their associated temporal values:
\begin{equation} \label{equ:qft_12}
s_{t_n}(n_A) = \left\{ \begin{array}{ll}
s_{e_n}(n=2 \cdot n_A) \quad &  n \in nze\_n(s_{e_n}) \\
0 \quad & \text{otherwise}  \\
\end{array} \right.
\end{equation}
From a frequency-domain perspective, in $\DCT$ context, the following relation holds:
\begin{equation} \label{equ:qft_13}
\DCT[s_{e_n}](k) = \DCT[s_{t_n}](k) \quad k \in sto\_k(s_{e_n})
\end{equation}
while in $\DST$ context the same eq.(\ref{equ:qft_13}) holds, changing $\DCT$ with $\DST$. 
Here we list the children signal types (described in Tab.\ref{tab:notation}) obtained applying this decomposition to different mother signal types.

If $s_{e_n}=s_{dc\_et}$ holds then $s_{t}=s_{dc\_tt}$.

If $s_{e_n}=s_{ds\_et}$ holds then $s_{t}=s_{ds\_tt}$.


\subsection{notes on the used basic elaborations} 

We highlight that the eq.(\ref{equ:qft_5A}),(\ref{equ:qft_5B}),(\ref{equ:qft_1}),(\ref{equ:qft_2}),(\ref{equ:qft_9}),(\ref{equ:qft_12}) prove that the basic elaborations used in this paper create descendent signals with many a-priori known $s(n)=0$ values, that thus don't require to be computed or stored.
Moreover the eq.(\ref{equ:qft_5C}),(\ref{equ:qft_5D}),(\ref{equ:qft_6}),(\ref{equ:qft_4B}),(\ref{equ:qft_10}),(\ref{equ:qft_13}) prove that we can compute $S(k)$ values of mother signals of each basic elaboration, computing (and storing) frequency-domain values just in a subset ($sto\_k$) of harmonics of descendent signals, created by each basic elaborations.
These observations legitimate the choise of creating new concepts, and a notation, that focus on $sto\_n$ and $sto\_k$ groupings of each created pruned input and/or output signal.

\section{The classical QFT algorithm} \label{sec:la_QFT_classica}

The QFT algorithm \cite{Guo_Sitton_qft_1994} (here denoted as classical QFT to distinguish it from the improved QFT algorithm), is a real-factor algorithm which has encountered a significant success. 
We can describe it in terms of six functions calling each other, if it is finalized to the computation of the $\CDFT$. 
Although it can be found in \cite{Guo_Sitton_qft_1994}, here we present it using the new terms (signal types, notation, basic elaborations) developed in the previous sections. 
In this way the differences with the new algorithm (proposed in the next section), as the reason of inefficiency of classical version, will be clearly evident.

\subsection{The $cdft\_cla$ function}

The input signal is of type $s_{cx\_tt}$. Let $N$ be its length (equal to periodization).
If $N=2$ then the $\CDFT$ definition is directly applied, otherwise the $\CDFT$ calculation is decomposed into two $\RDFT$ tranforms, relative to children output signals $s_{1\_re\_tt}$ and $s_{2\_re\_tt}$, both of length $N$ (equal to periodization) that we manage by means of $rdft\_cla$ function.
The two children signals are created according to the following time domain relations:
\begin{align*} 
& s_{1\_re\_ tt}(n)= \Re[s_{cx\_ tt}(n)] \quad n \in \{0,1,2,\dots,(N-1) \}  \\
& s_{2\_re\_ tt}(n)= \Im[s_{cx\_ tt}(n)] \quad n \in \{0,1,2,\dots,(N-1) \} 
\end{align*}
We can prove that above time domain relations correspond to the following frequency-domain relationships (backward phase):
\begin{equation*}
\begin{split} \label{equ:qft_17}   
\Re \{\CDFT[ s_{cx\_tt}]\}(k) & = \Re \{\RDFT[s_{1\_re\_ tt}]\}(k) - \Im \{\RDFT[s_{2\_re\_tt}] \}(k) \\
& \quad k \in \{1,2,\dots,(\frac{N}{2}-1)\} 
\end{split} \\
\end{equation*}
\begin{equation*}
\begin{split} \label{equ:qft_18}
\Re \{\CDFT[ s_{cx\_tt}]\}(N-k) & = \Re \{\RDFT[s_{1\_re\_tt}]\}(k) + \Im \{\RDFT[s_{2\_re\_tt}] \}(k)  \\
& \quad  k \in  \{1,2,\dots,(\frac{N}{2}-1)\} 
\end{split} \\
\end{equation*}
\begin{equation*}
\Re \{\CDFT[ s_{cx\_tt}]\}(k)= \Re \{\RDFT[s_{1\_re\_tt}]\}(k) \quad k \in \{ 0, (\frac{N}{2}) \} \label{equ:qft_19}
\end{equation*}
\begin{equation*}  
\begin{split} \label{equ:qft_19A}
 \Im \{\CDFT[ s_{cx\_tt}]\}(k) & = \Im \{\RDFT[s_{1\_re\_tt}]\}(k) + \Re\{\RDFT[s_{2\_re\_tt}] \}(k) \\
& \quad k \in \{1,2,\dots,(\frac{N}{2}-1)\}  
\end{split} \\
\end{equation*}
\begin{equation*}
\begin{split} \label{equ:qft_20}
 \Im \{\CDFT[ s_{cx\_tt}]\}(N-k) & = -\Im \{\RDFT[s_{1\_re\_tt}]\}(k) + \Re \{\RDFT[s_{2\_re\_tt}] \}(k) \\
& \quad k \in  \{1,2,\dots,(\frac{N}{2}-1)\}
\end{split} \\
\end{equation*}
\begin{equation*}
  \Im \{\CDFT[ s_{cx\_tt}]\}(k) = \Re \{\RDFT[s_{2\_re\_tt}]\}(k) \quad k \in \{ 0, (\frac{N}{2}) \} \label{equ:qft_21}
\end{equation*}

\subsection{The $rdft\_cla$ function in classical QFT}

The input signal is of type $s_{re\_tt}$. Let $N$ be its length (equal to periodization).
If $N=2$ then we apply the $\RDFT$ definition, otherwise we decompose the $\RDFT$ calculation into the calculation of a $\DCT$ (applied to the child signal $s_{dc\_tt}$ of periodization $N$, that we manage through a $dct\_cla$ function) and a $\DST$ (applied to the child signal $s_{ds\_tt}$ of periodization $N$, that we manage through a $dst\_cla$ function). We can prove that the two output time domain children signals are created by means of these equations:
\begin{equation*} \label{equ:qft_22}
s_{dc\_tt}(n)= \left\{ \begin{array}{ll}
s_{re\_tt}(n) + s_{re\_tt}(N-n) \quad & n \in \{1,2,3,\dots,(\frac{N}{2}-1) \} \\
s_{re\_tt}(n) \quad & n \in \{0, (\frac{N}{2}) \} \\
0 \quad & \text{otherwise}  \\
\end{array} \right.
\end{equation*}
\begin{equation*} \label{equ:qft_23}
s_{ds\_tt}(n)= \left\{ \begin{array}{ll}
s_{re\_tt}(n) - s_{re\_tt}(N-n) \quad & n \in \{1,2,3,\dots,(\frac{N}{2}-1) \} \\
0 \quad & \text{otherwise}  \\
\end{array} \right.
\end{equation*}
corresponding to the following frequency-domain relationships (backward phase):
\begin{align*}
& \Re \{\RDFT[s_{re\_tt}]\}(k) = \DCT[s_{dc\_tt}](k) \quad & k \in \{ 0,1,2,\dots,(\frac{N}{2}) \}  \\
& \Im \{\RDFT[s_{re\_tt}]\}(k) = - \DST[s_{ds\_tt}](k) \quad & k \in \{ 1,2,3,\dots,(\frac{N}{2}-1) \} 
\end{align*}

\subsection{The $dct\_cla$ function} \label{dct_function}

The input signal is of type $s_{dc\_tt}$. 
Let $N$ be its periodization.
If $N=2$ then we directy apply $\DCT$ definition, otherwise the mother signal $s_{dc\_tt}$ is first decomposed separating the even and odd harmonics (as shown in sect. \ref{sec:separe_k}), creating the two children signals $s_{dc\_te}$ and $s_{dc\_to}$, both with periodization equal to $N$. 
Afterwards, the signal $s_{dc\_te}$ is converted into the signal $s_{A\_dc\_tt}$, having periodization $N_A=(\frac{N}{2})$, by halving each even $k$ index (as shown in sec. \ref{sec:k_even}). 
Therefore we obtain the two output signals $s_{A\_dc \_tt}$ and $s_{dc\_to}$, that we handle by the $dct\_cla$ and $dct\_to\_cla$ functions respectively.

\subsection{The $dst\_cla$ function}

This function operates similarly to the $dct$ one.
Moreover the input and output signals have the same notation as in the $dct$ case, a part from substituting $dc$ with $ds$ (this is possible thanks to our notation, 
which highlights the analogy between $\DCT$ and $\DST$ management of signals).

\subsection{The $dct\_to\_cla$ function}

The input signal is of type $s_{dc\_to}$. Let's call $N$ its periodization.
If $N=4$ then we appy the $\DCT$ definition, otherwise we apply the following operations.
First, we transform the sequence of odd harmonics of mother signal $s_{dc\_to}$, into the sequence of even harmonics of a new signal, of type $s_{dc\_t_{1} e}$, by multiplying the signal $s_{dc\_to}$ with the secant function in the time domain:
\begin{equation} \label{equ:qft_26}
s_{dc\_t_{1} e}(n)= s_{dc\_to}(n) \cdot \frac{1}{2 \cdot \cos(\theta \cdot n)} \quad n \in \{0,1,2,\dots,(\frac{N}{4}-1) \}
\end{equation}
In eq.(\ref{equ:qft_26}) a special case holds for $n=0$, since the multiplication can be substituted by a binary translation.
It can be proved that eq.(\ref{equ:qft_26}) corresponds to the following frequency-domain relationship (backward phase):
\begin{equation}
\begin{split} \label{equ:qft_28}
\DCT[s_{dc\_to}](k) &= \DCT[s_{dc \_t_{1} e}](k-1) + \DCT[s_{dc \_t_{1} e}](k+1) \\
& \quad k \in \{ 1,3,5,\dots, (\frac{N}{2}-1) \}
\end{split}
\end{equation}
Afterwards, we halve the even $k$ indices of signal $s_{dc\_t_{1} e}$ transforming it into the signal $s_{A \_dc t_{1}t}$, which have periodization $N_A=(\frac{N}{2})$ (as shown in sect. \ref{sec:k_even}). 
We then process this signal similarly to $s_{dc\_tt}$ in the $dct$ function (sect. \ref{dct_function}). 
As a result we have two output signals $s_{B \_dc \_tt}$ (with periodization $N_B=(\frac{N}{4})$) and $s_{A \_dc \_to}$ (with periodization $N_A=(\frac{N}{2})$), that we handle by the $dct\_cla$ and $dct\_to\_cla$ functions respectively.

\subsection{The $dst\_to\_cla$ function}

This function is different from the $dct\_to$ one in many aspects since it applies to the input signal $s_{ds\_to}$. 
We first separate the $nze\_n(s_{ds\_to})$ indices into two children signals: $s_{ds\_t_{1} o}$ and $s_{ds\_e_{1}o}$ ($s_{ds\_e_{1}o}$ ereditates, and has, only 
the residual time index $n=(\frac{N}{4})$):
\begin{equation} \label{equ:qft_29}
 s_{ds\_t_{1} o}(n) =
\begin{cases}
s_{ds\_to}(n) \quad & n \in \{ 1,2,3,\dots,(\frac{N}{4}-1) \}    \\
  0  \quad & n=\frac{N}{4} \\
  0  \quad & n \in \{ 0,(\frac{N}{4}+1),(\frac{N}{4}+2), \dots, (N-1) \} \\
\end{cases}
\end{equation} 
\begin{equation} \label{equ:qft_29b}
s_{ds\_e_{1} o}(n) =
\begin{cases}
s_{ds\_to}(n) \quad & n= \{ \frac{N}{4} \}    \\
 0  \quad & n \in \{ 0,1,\dots,(\frac{N}{4}-1) \}  \\
\end{cases}
\end{equation}
We can prove that eq.(\ref{equ:qft_29}), (\ref{equ:qft_29b}) correspond to the following frequency-domain relationship (backward phase):
\begin{equation} \label{equ:qft_29c}
\begin{split}
\DST[s_{ds\_to}](k) &= \DST[s_{ds\_t_{1} o}](k)+(-1)^\frac{k-1}{2} \cdot \DST[s_{ds\_e_{1} o}](k=1) \\
& \quad k \in sto\_k(s_{ds\_to})
\end{split}
\end{equation}
where:
\begin{equation} \label{equ:qft_29d}
\DST[s_{ds\_e_{1} o}](k=1)=s_{ds\_e_{1} o}(n=\frac{N}{4})
\end{equation}
Then we transform $s_{ds\_t_{1} o}$ into $s_{ds\_te}$, as done in (\ref{equ:qft_26}),(\ref{equ:qft_28}) (only involved n\_indeces and signal types change), and then we transform $s_{ds\_te}$ into $s_{A\_ds\_tt}$ by halving each $k$ index (as described in sect. \ref{sec:k_even}).
Differently from $dct\_to\_cla$ function, we don't prosecute applying to $s_{A\_ds\_tt}$ the separation between even and odd harmonics (described in sect. \ref{sec:separe_k}), since we have already created two output signals in this function.
Thus the two output signals are: $s_{A\_ds\_tt}$ (handled by $dst\_cla$ function) and $s_{ds\_e_1 o}$ that is a leaf of decomposition tree (see 
Fig.\ref{fig:classical_qft}) and, for this reason, it is not handled by other functions.
 
Let us observe that the creation of $s_{ds\_e_{1} o}$ signal means to compute separately (using the definition of $\DST$, a very inefficient tecnique) the contribution of $s_{ds\_to}(n=\frac{N}{4})$ temporal element to frequency-domain components of $s_{ds\_to}$, since we cannot apply (\ref{equ:qft_26}) to $(n=\frac{N}{4})$ case too (to avoid division by zero).


\subsection{The classical QFT: the decomposition tree}

Classical QFT recursive algorithm can be diagrammatically represented by the decomposition tree reported in Fig.\ref{fig:classical_qft}.
Such a tree refers to calculating the entire $\RDFT$ of a $s_{re\_tt}$ root signal, and reports four levels of decomposition (the total number of decomposition levels depends on the signal periodization $N$). 
In order to facilitate the identification of different signals within the decomposition tree, which could be difficult due to the occurrence of multiple usage of the same functions and of the same signal types many times, each signal has been re-named (with respect to the names used in this section), according both to sect. \ref{sec:notation_sub_2} and to the following criterion: a $n_1\_n_2$ identifier is placed in front of the subtitles symbols list, where $n_1$ denotes the decomposition level, and $n_2$ the signal position within the decomposition level.
It follows that the starting signal, which is processed by the function $rdft$, changes its name from $s_{re\_tt}$ to $s_{1,1\_re\_tt}$. 
Moreover the two output signals created by this function become $s_{2,1\_dc\_tt}$ and $s_{2,2\_ds\_tt}$. 
They are respectively processed by functions $dct$ and $dst$. 
The $dct$ function (whose intermediate signals created inside it are also reported in Fig.\ref{fig:classical_qft}) generates the two output signals $s_{A\_dc\_tt}$  ($s_{3,1 A\_dc\_tt}$) and $s_{dc\_to}$ ($s_{3,2\_dc\_to}$), which are the input signals of functions $dct\_cla$ and $dct\_to\_cla$ respectively. 
The remaining part of the graph can be explained in a similar way.


\subsection{The classical QFT: computational cost}

For $\CDFT$ calculation, the classical QFT presents the following computational cost \cite{Guo_Sitton_qft_1994}:
\begin{align*}
& \mul(N) =  N \cdot \log(N) - \frac{11}{4} \cdot N + 2 \\
& \add(N) = \frac{7}{2} \cdot N \cdot \log(N) - 4 \cdot N  \\
& \flop(N) =  \frac{9}{2} \cdot N \cdot \log(N) - \frac{27}{4} \cdot N + 2  
\end{align*}

\subsection{The classical QFT: memory requirements} \label{sec:memory_old_qft}

The classical QFT algorithm requires the employment of $(\frac{N}{4}-1)$ distinct trigonometric (secant) constants: $\csc(\theta \cdot n) \quad \text{for} \quad n \in \{ 0,1,\dots,(\frac{N}{4}-1)\}$.
Moreover the classical QFT can be implemented in-place only if our goal is the $\DST-0$ computation (an implementation is in-place only if the memory size needed to perform the algorithm operations, in addition to the memory area containing the start-signal, is fixed, that means not dependent on the periodization $N$ of handled root signal).
In fact each function used in $\DST$ context (intuitively speacking) has these good characteristics:
\begin{itemize}

\item 
it doesn't increase total $ln$ and total $lk$ to handle, passing from the input mother signal to the two output descendent signals.

\item 
it can be implememnted using a fixed number of inner temporary variables, that doesn't depend on the periodization $N$ of handled input signal.

\item
it uses basic elaborations whose conversions or combinations of temporal (or frequency-domain) elements can intrinsically be implemented in-place (not depending on algorithm where they are inserted).

\end{itemize}
Differently the computation of $\DCT-0$ or $\DFT$, can not be implemented in-place, using the classical QFT, because of $dct\_to\_cla$ function, that increases both the total $sto\_n$ indices and total $sto\_k$ harmonics that we have to handle (and to store in memory), passing from the input signal to the two output descendent signals, each time this function is used.
Here is the proof of this statement.
From  Tab.\ref{tab:notation}, considering the specific periodization of each signal, the following relations hold 
(from an input-output point of view), for the $dct\_to\_cla$ function:
\begin{align*}
&ln(s_{dc\_to})=\frac{N}{4} &lk(s_{dc\_to})=\frac{N}{4}  \\
&ln(s_{A\_dc\_to})=\frac{N_A}{4}=\frac{N}{8} &lk(s_{A\_dc\_to})=\frac{N_A}{4}=\frac{N}{8}  \\
&ln(s_{B\_dc\_tt})=\frac{N_B}{2}+1=\frac{N}{8}+1  &lk(s_{B\_dc\_tt})=\frac{N_B}{2}+1=\frac{N}{8}+1 
\end{align*}
Combining the previous values we obtain the thesis in the $dct\_to\_cla$ case:
\begin{align}
&ln(s_{dc\_to})=ln(s_{A\_dc\_to})+ln(s_{B\_dc\_tt})+1 \label{equ:qft_37} \\
&lk(s_{dc\_to})=lk(s_{A\_dc\_to})+lk(s_{B\_dc\_tt})+1 \label{equ:qft_38}
\end{align}
This implies an increment of total elements to manage as progressing the tree decomposition, and therefore high memory requirements
are needed (if $N$ is high), preventing an `in place' implementation too.
The shown formal proof has a disadvantage: it doesn't explain the mechanism inside the $dct\_to\_cla$ function that creates eq.(\ref{equ:qft_37}),(\ref{equ:qft_38}). 
Here is the explanation of this mechanism.
Total $lk$ increases because of (\ref{equ:qft_28}) that forces us to know (and to store) $\frac{N}{4}$ harmonics of child signal $s_{dc\_t1e}$, to compute only $\frac{N}{4}-1$ harmonics of mother signal $s_{dc\_to}$, in backward phase.
On the contrary eq.(\ref{equ:qft_26}) doesn't increase $ln$.
For this reason (\ref{equ:qft_26}),(\ref{equ:qft_28}) create a child signal ($s_{dc\_t1e}$) with $lk>ln$ (and thus with $ind\_sto\_k \neq sto\_k$ too).
Handling a signal ($s_{dc\_t1e}$) with $lk=ln+1$ is inefficient because it means we have to compute and to store in memory an harmonic $k$ whose $S(k)$ value is linearly dependent from all other ones of the signal, since a pruned input signal with only $ln$ real independent temporal elements can have a maximum of $ln$ independent real transformed elements (thus, intuitively speaking, this adding harmonic doesn't increase the information stored in other harmonics).
The increase of total $ln$ inside $dct\_to\_cla$ function is caused by the separation between even and odd harmonics of $s_{dc\_t1t}$ signal, used after (\ref{equ:qft_26}),(\ref{equ:qft_28})).
This theoretical elaboration increases total $ln$ because we apply it to an atipical signal ($s_{dc\_t1t}$) that has $ln \neq lk$ (and thus with $ind\_sto\_k \neq sto\_k$ too).
In fact this problematic is not intrinsic of the decomposition described in sect. \ref{sec:separe_k} since, applying this decomposition to other signal types, total $ln$ and $lk$ parameter are kept inalterated.


\section{The improved QFT algorithm: basic idea and description} \label{sec:improved_QFT}

\subsection{The idea of improved QFT algorithm}

In classical QFT algorithm, two are the main factors that make the computational cost high: 
\begin{itemize}

\item
the separate handling  of time domain element $s_{ds\_to}(n=\frac{N}{4})$, within the $\DST$ calculation (we compute its contribution to $\DST$ frequency-domain components by means of $\DST$ definition to avoid the division by zero required if we apply (\ref{equ:qft_26}) to $s_{ds\_to}(n=\frac{N}{4})$).

\item
the increasing global number of elements to be managed, both temporal and frequency-domain, in $\DCT$ context, as signal decomposition proceeds further, because of $dct\_to\_cla$ function, as shown in sect. \ref{sec:memory_old_qft}.

\end{itemize}
The idea to improve this algorithm consists in applying both the separection between even and odd n-indices and the separation between even and odd harmonics before to transform odd in even k-indices. 
In this way we avoid the growth of both temporal and frequeny-domain elements to be managed as the decomposition level raises (in $\DCT$ context), as we will prove in sect. \ref{sec:memory_new_qft}.
Moreover, in $\DST$ context, the odd in even harmonics conversion will affect signals with only odd $n$ indices, avoiding to handle the problematic even index $n=\frac{N}{4}$, and thus avoiding to compute its contribution to frequency-domain components using the definition of $\DST$ (a very inefficient tecnique used in $dst\_to$ function in classical QFT). 

Let us stress that the separation betweeen even and odd time indeces can be performed both before and after the even/odd harmonics separation. 
We prefer to apply the temporal separation first, and then the frequency-domain one, in order to minimize the number of distinct recursive functions to involve. 
According to these modifications in the new QFT algorithm, the odd in even conversion is applied only to $s_{dc\_oo}$ and $s_{ds\_oo}$ signal types.


\subsection{Recursive description of improved QFT  algorithm} \label{improved_qft}

Improved QFT algorithm can be described in terms of 8 functions ($cdft$, $rdft$, $dct$, $dst$, $dct\_ot$, $dst\_ot$, $dct\_oo$, $dst\_oo$) calling each other, if it is finalized to the computation of $\CDFT$.
The $rdft$ and $cdft$ functions coincide with $rdft\_cla$ and $cdft\_cla$ functions respectively used in classical QFT algorithm.

\subsubsection{The $dct$ function}

The input signal is of type $s_{dc\_tt}$, with periodization $N$.
If $N=2$ we just apply the $\DCT$ definition, otherwise we first separate the even temporal indices from the odd ones (as shown in sect. \ref{sec:separe_n}), generating the signals $s_{ds\_et}$ and $s_{dc\_ot}$, characterized by periodization $N$.
Then we convert the signal $s_{ds\_et}$ into the signal $s_{A \_dc tt}$, with periodization $N_A=(\frac{N}{2})$, by halving each even temporal index $n$ (as shown in sect. \ref{sec:n_even}). 
At the end we have the two output signals $s_{A\_dc\_tt}$ and $s_{dc\_ot}$ handled by, respectively, the $dct$ and the $dct\_ot$ function.
Let us observe that this function (as $dst$ function) contain the new decomposition (the separation between dd and even time indeces) introduced in improved QFT algorithm to avoid the problematics of classical QFT.

\subsubsection{The $dct\_ot$ function}

This function operates as the $dct\_cla$ function in classical QFT, but it applies to $s_{dc\_ot}$ signal type (with periodization $N$), instead of  $s_{dc\_tt}$  signal type.
If $N=4$  we just apply the $\DCT$ definition, otherwise we first separate even from odd harmonics (as shown in sect. \ref{sec:separe_k}), generating the signals $s_{dc\_oe}$ e $s_{dc\_oo}$, both characterized by periodization $N$.
Then the signal $s_{dc\_oe}$ is converted into the signal $s_{A \_dc \_ot}$, having periodization $N_A=(\frac{N}{2})$, by halving each even index $k$ (as shown in sect. \ref{sec:k_even}). 
We therefore obtain two output signals $s_{A\_dc\_ot}$ and $s_{dc\_oo}$ handled by, respectively, the $dct\_ot$ and $dct\_oo$ functions.

\subsubsection{The $dct\_oo$ function}

This function operates as the $dct\_to\_cla$ function in classical QFT, but it applies to $s_{dc \_oo}$ signal type, instead of $s_{dc \_to}$ signal type, being this slight difference relevant to improve the $\DCT$ computational cost. 
The input signal is of type $s_{dc\_oo}$, with periodization $N$.
If $N=8$ we just apply the $\DCT$ definition, otherwise we first transform the even harmonics into the odd ones, generating the child signal $s_{dc \_oe}$, with periodization $N$, by means of these equations:
\begin{equation} \label{equ:qft_38A}
s_{dc\_oe}(n)= s_{dc\_oo}(n) \cdot \frac{1}{2 \cdot \cos(\theta \cdot n)} \quad n \in \{0,1,2,\dots,(\frac{N}{4}-1) \}
\end{equation}
\begin{equation}
\DCT[s_{dc \_oo}](k=\frac{N}{4}-1) = \DCT[s_{dc \_oe}](k=\frac{N}{4}) \label{equ:qft_38B} 
\end{equation}
\begin{equation}
\begin{split} \label{equ:qft_38C}
\DCT[s_{dc\_oo}](k) &= \DCT[s_{dc \_oe}](k-1) + \DCT[s_{dc \_oe}](k+1) \\
& \quad k \in \{ 1,3,5,\dots, (\frac{N}{4}-1) \}
\end{split}
\end{equation}
These relations are similar to eq.(\ref{equ:qft_26}),(\ref{equ:qft_28}), the only difference being involved signal types. 
Then we halve the even $k$ indices of $s_{dc\_oe}$ (as shown in sect. \ref{sec:k_even}), generating the signal  $s_{A \_dc \_ot}$, having periodization $N_A=(\frac{N}{2})$.
From now on we handle the signal $s_{A \_dc \_ot}$  as well as we handle the signal $s_{dc\_ot}$ in function $dct\_ot$ of this algorithm, and therefore we create two output signals $s_{B \_dc \_ot}$ (with periodization $N_B=(\frac{N}{4})$, handled by function $dct\_ot$) and $s_{A \_dc \_oo}$ (with periodization $N_A=(\frac{N}{2})$, handled by function $dct\_oo$).

\subsubsection{The $dst$, $dst\_ot$, $dst\_oo$ functions} \label{sec:new_qft_dst_functions}

The functions $dst$, $dst\_ot$, $dst\_oo$ apply the same chain of elaborations used in $dct$, $dct\_ot$, $dct\_oo$ functions respectively, but applied to different input signal types ($s_{ds\_tt}$, $s_{ds\_ot}$, $s_{ds\_oo}$ instead of $s_{dc\_tt}$, $s_{dc\_ot}$, $s_{dc\_oo}$ respectively).
Thus the signal types used in this $dst$ family of functions have the same notation of the
corresponding ones used in $dct$ family, the only difference being that each $dc$ occurance is
replaced by the $ds$ one. 
Moreover, two other relevant differences occur:
\begin{itemize}
\item 
in $dst$ function the $\DST$ definition is applied for $N=4$ instead of $N=2$.

\item
in $dst\_oo$ function, a special case holds for $k=1$, instead of  $k=(\frac{N}{4}-1)$, in eq.(\ref{equ:qft_38B}).

\end{itemize}
Improved QFT recursive algorithm can be diagrammatically represented by the decomposition tree reported in Fig.\ref{fig:improved_qft}.


\section{characteristics of the improved QFT algorithm} \label{sec:results}

In this section we compare the characteristics of improved QFT with the ones of classical QFT and of split-radix 3mul-3add


\subsection{Memory Requirements} \label{sec:memory_new_qft}

\begin{itemize}
\item 
Improved QFT algorithm requires few trigonometric constants: quite the same used in classical QFT.

In fact it employs $(\frac{N}{4})$ distinct trigonometric constants: the same $(\frac{N}{4}-1)$ secant constants used in classical QFT algorithm, more the $\cos(\frac{2 \cdot \pi}{8})$ constant used in $dct\_oo$ and $dst\_oo$ function if $N=8$ (the special case where the definitions of $\DCT$ and $\DST$ respectively are applied).
It is a good characteristic since, for example, conjugate-pair split-radix 3add-3mul requires twice real trigonometric constants.

\item
Improved QFT algorithm requires less memory cells than classical QFT, if the goal is the $\DCT$ or $\DFT$ computation.

The reason is that, in any recursive function (or in any level of decomposition), of the new QFT algorithm, the total $ln$ or $lk$ parameters do not increase, passing from the input to the descendent output signals (differently these parameters increase in $\DCT$ context, in classical QFT, as seen in sect. \ref{sec:memory_old_qft}).

We prove this statement here only for the $dct\_oo$ function, since it plays the same role that the `ill' $dct\_to\_cla$ function (the one that increases total $ln$ and $lk$) has in classical algorithm (since both functions contain the conversion of odd harmonics signal into an even harmonics signal, multiplying the mother signal by secant function).

From Tab.\ref{tab:notation}, considering the
specific periodization of each created signal, the following relations hold (from an input-output point of view) for $dct\_oo$ function:
\begin{align*}
&ln(s_{dc\_oo})=\frac{N}{8} &lk(s_{dc\_oo})=\frac{N}{8}  \\
&ln(s_{A\_dc\_oo})=\frac{N_A}{8}=\frac{N}{16} &lk(s_{A\_dc\_oo})=\frac{N_A}{8}=\frac{N}{16}  \\
&ln(s_{B\_dc\_ot})=\frac{N_B}{4}=\frac{N}{16} &lk(s_{B\_dc\_ot})=\frac{N_B}{4}=\frac{N}{16} 
\end{align*}
Combining the previous values we obtain the thesis (for $dct\_oo$ case):
\begin{align}
&ln(s_{dc\_oo})=ln(s_{A\_dc\_ot})+ln(s_{B\_dc\_ot}) \label{equ:qft_42} \\
&lk(s_{dc\_oo})=lk(s_{A\_dc\_ot})+lk(s_{B\_dc\_ot}) \label{equ:qft_43}
\end{align}
Similar relations hold for any other function used in improved QFT algorithm.
Comparing (\ref{equ:qft_42}),(\ref{equ:qft_43}) with (\ref{equ:qft_37}),(\ref{equ:qft_38}), it results that the new algorithm requires
less memory locations than classical QFT, in $\DCT$, and thus $\DFT$ too, context.

Let us analize the mechanism that avoid us to obtain eq.(\ref{equ:qft_37}),(\ref{equ:qft_38}) in improved QFT.
In classical and in improved QFT we convert odd harmonics signal into an even harmonics signal, in the same manner: multiplying temporal signal by secant function (only involved signal types change). 
However, applying this theoretical elaboration to $s_{dc\_oo}$  we obtain a child signal ($s_{dc\_oe}$) with $ln=lk$.
Differently in classical QFT applying the same conversion to $s_{dc\_to}$ signal we create the $s_{dc\_t1e}$ signal type, that has $lk>ln$.
Moreover, in improved QFT, applying the separation between even and odd harmonics (described in sect. \ref{sec:separe_k}) to $s_{dc\_ot}$ (instead of $s_{dc\_t1t}$ used in classical QFT) total $ln$ doesn't increase. 
For these reasons all signal types created in improved algorithm have $ln=lk$ (and $ind\_sto\_k=sto\_k$ too), and thus we remove the inefficiences described in sect. \ref{sec:memory_old_qft}.
We highlight once again that the shown slight differences of improved QFT versus classical QFT are difficult to cath using traditional exposition, but are easy to catch using new exposition approach here used, that focus on $sto\_n$ and $sto\_k$ groupings of each created signal.

\item
The new QFT algorithm is eligible for an `in place' implementation (but to find an efficient code that implements it requires further work), because each used function has the characteristics already described in sect. \ref{sec:memory_old_qft} for the functions used in $\DST$ context, in classical QFT.

\end{itemize}


\subsection{Computational Cost} \label{sec:costo_computazionale}

The mathematical expressions of detailed computational cost of the new algorithm are reported in Tab.\ref{tab:costo_improved_qft}. 
Tab.\ref{tab:compara_add}, \ref{tab:compara_mol}, \ref{tab:compara_flop} compare the computational cost of improved QFT with those of classical QFT and of split-radix 3add-3mul \cite{Sorensen_1986} in the $\CDFT$ case.
It results that the improved QFT:
\begin{itemize}

\item
requires less additions, multiplications, and flops (and, for this reason, is more efficient, in this model of computation), than the classical QFT, for $N \geq 16$. 

\item 
requires the same additions, multiplications and flops of split-radix 3add/3mul for any $N=2^r$, but using half trigonometric constants, as seen before and according to \cite{Johnson_Frigo_2007},\cite{Sorensen_1986}.

It can be shown that improvements are more consistent in the $\DST$ than in the $\DCT$ case, since, in the former case, we avoid to use eq. (\ref{equ:qft_29c}). 
Intuitively speaking, the improved QFT algorithm has a lower computational cost than classical QFT for the same reasons shown in sect. \ref{sec:memory_new_qft} for memory requirements, since computational cost is linearly related to $ln$ and $lk$ parameters of handled signals. 
In fact more indeces we handle, more arithmetic intructions we compute (to store indeces that will not be used again, has no sense).

\end{itemize}



\subsection{numerical accuracy} \label{sec:accuracy}

\begin{figure}[htb]
  \centering
  \includegraphics[width=0.7 \textwidth, angle=0]{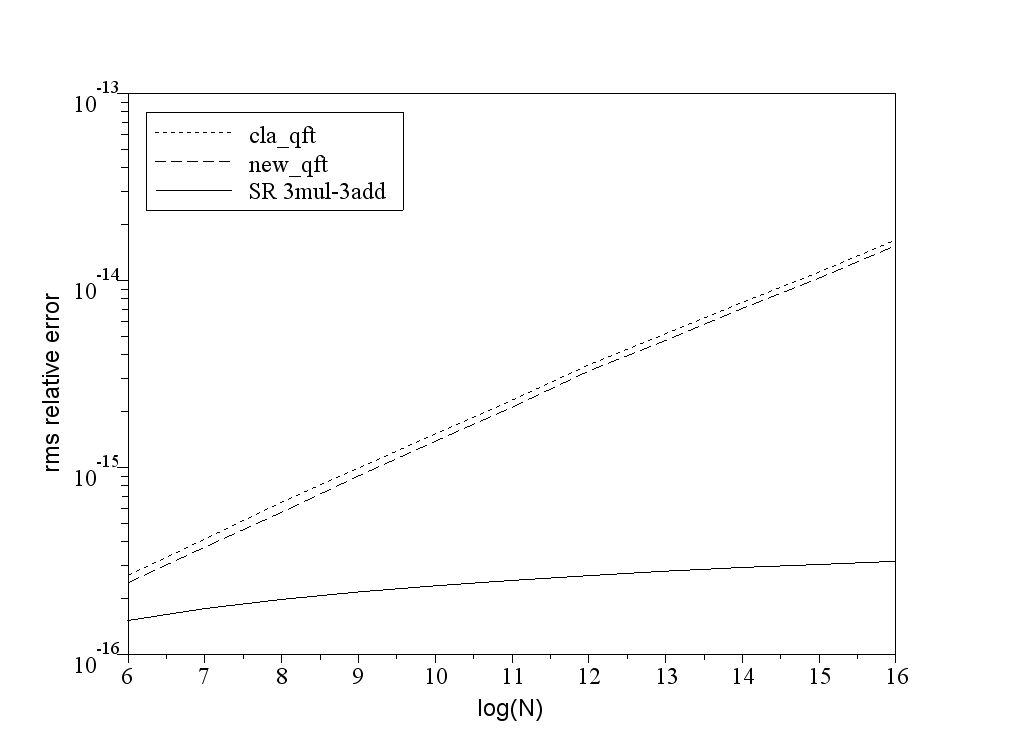}
  \caption{Comparison in numerical accuracy of improved QFT (new\_QFT), classical QFT (cla\_QFT) and split-radix (SR) 3mul-3add}
  \label{fig:accuracy}
\end{figure}

We have tested and compared  (see Fig.\ref{fig:accuracy}) the accuracy of improved QFT, classical QFT and split-radix 3add-3mul, using Scilab 5.3.3 and 64 bit double precision data types, on a Pentium IV with MS Windows XP.
The accuracy of each algorithm has been quantified by means of relative rms error (according to \cite{Johnson_Frigo_2007}) and testing the algorithms on many ($10^{2}-10^{3}$ depending on $N$) $h$ random (with $-0.5<|\Re[s_i(n)]|<0.5, \quad -0.5<|\Im[s_i(n)]|<0.5 \quad \forall n$) complex-value signals $s_i$:
\begin{equation} \label{equ:qft_44}
relative\_rms\_error=\frac{1}{h}\sum_{i=0}^{h-1} \frac{|| S_i-S\_exact\_i ||}{|| S\_exact\_i ||}
\end{equation}
In (\ref{equ:qft_44}) the euclidean norm is used, $S\_exact\_i=\text{exact-CDFT}[s_i]$ is computed using quadruple precision, $S_i=\text{approx-CDFT}[s_i]$ is the estimated output signal obtained using the handled FFT algorithm and double precision computation.  
We highlight that, in $S_i=\text{approx-CDFT}[s_i]$ computation, any used double-precision trigonometric constant has been pre-computed passing through a quadruple-precision value, and then rounding it in double-precision.
In this manner both the used trigonometric constants array and the tested FFT algorithm have the best accuracy with respect the limit of 64 bit storage (this optimal accuracy can not be directly reached using the sine and cosine 64-bit default functions). 
It is an interesting aspect since, if we would compute the required trigonometric constants remaining in double precision (without passing through quadruple precision), then the numerical error of  classical and improved algorithms will grow about 22\% and 25\% respectively for $N=2^{16}$.
Fig.\ref{fig:accuracy} shows that improved QFT enhances the accuracy of classical QFT about 4-9\% (depending on $N$).
Moreover for small $N$ the accuracies of improved QFT and of split-radix 3add-3mul are quite the same. 
Unfortunately both classical and improved QFT have a numerical error that grows faster than that one of split-radix 3mul-3add.
However we think that it often is not a so relevant disadvantage.
In fact, in many applications, we are interested only on the first three decimal digits of the output values $S(k)$ and, in such a context, to have a $relative\_rms\_error$ equal to $10^{-14}$ or $10^{-16}$ is quite the same. 
Thus we can use the improved QFT also if $N$ is high in such applications.

\begin{table}[tb]
\caption{Computational cost required for various sinusoidal transforms by means of the improved QFT, in dependence on their periodization $N$.} 
\label{tab:costo_improved_qft}
\centering
\scalebox{0.8}
{
\begin{tabular}{cccc}
\toprule
\cmidrule{2-4}
transform & multiplications & sums & flop \\
\midrule
CDFT & $N \log(N)-3 N+4$ & $3 N \log(N)-3 N+4$ & $4 N \log(N)-6 N+8$ \\
RDFT & $\frac{1}{2} N \log(N)- \frac{3}{2} N+2$ & $\frac{3}{2} N \log(N)-\frac{5}{2} N+4$ & $2 N \log(N)-4 N+6$  \\
DCT-0 & $\frac{1}{4} N \log(N)- \frac{3}{4} N+1$ & $\frac{3}{4} N \log(N)-\frac{7}{4} N+\log(N)+3$ & $N \log(N)- \frac{5}{2} N+\log(N)+4$ \\
DST-0 & $\frac{1}{4} N \log(N)- \frac{3}{4} N+1$ & $\frac{3}{4} N \log(N)-\frac{7}{4} N-\log(N)+3$  & $N \log(N)- \frac{5}{2} N-\log(N)+4$ \\
\bottomrule
\end{tabular}
} 
\end{table}



\begin{table}[tb]
\caption{Comparative evaluation of number of sums required for $CDFT$ calculation for split-radix 3add/3mul algorithm (SR\_3/3), classical QFT algorithm 
(clas\_QFT), improved QFT algorithm (new\_QFT)}
\label{tab:compara_add}
\centering
\scalebox{0.7}
{
\begin{tabular}{cccc}
\toprule
N &  $SR\_3/3$ & $clas\_QFT$ & $new\_QFT$ \\
\midrule
$4$ & $16$ & $16$ & $16$ \\
$8$ & $52$ & $52$ &  $52$ \\
$16$ & $148$ & $160$ &  $148$ \\
$32$ & $388$ & $432$ & $388$ \\
$64$ & $964$ & $1088$ & $964$ \\
$128$ & $2308$ & $2624$ & $2308$ \\
$256$ & $5380$ & $6144$ & $5380$ \\
$512$ & $12292$ & $14080$ & $12290$ \\
$1024$ & $27652$ & $31744$ & $27652$ \\
$2048$ & $61444$ & $70656$ & $61444$ \\
\bottomrule
\end{tabular}
}
\end{table}



\begin{table}[tb]
\caption{Comparative evaluation of number of multiplications required for $CDFT$ calculation for split-radix 3add/3mul algorithm (SR\_3/3), classical QFT algorithm 
(clas\_QFT), improved QFT algorithm (new\_QFT) }
\label{tab:compara_mol}
\centering
\scalebox{0.7}
{
\begin{tabular}{cccc}
\toprule
N & $clas\_QFT$ & $new\_QFT$ & $SR\_3/3$ \\
\midrule
$4$ & $0$ & $0$ & $0$ \\
$8$ & $4$ & $4$ & $4$ \\
$16$ & $22$ & $20$ & $20$ \\
$32$ & $74$ & $68$ & $68$ \\
$64$ & $210$ & $196$ & $196$ \\
$128$ & $546$ & $516$ & $516$ \\
$256$ & $1346$ & $1284$ & $1284$ \\
$512$ & $3202$ & $3076$ & $3076$ \\
$1024$ & $7426$ & $7172$ & $7172$ \\
$2048$ & $16898$ & $16388$ & $16388$ \\
\bottomrule
\end{tabular}
}
\end{table}



\begin{table}[tb]
\caption{Comparative evaluation of number of flops required for $CDFT$ calculation for split-radix 3add/3mul algorithm (SR\_3/3), classical QFT algorithm 
(clas\_QFT), improved QFT algorithm (new\_QFT)}
\label{tab:compara_flop}
\centering
\scalebox{0.7}
{
\begin{tabular}{cccc}
\toprule
N & $SR\_3/3$ & $clas\_QFT$ & $new\_QFT$ \\
\midrule
$4$ & $16$ & $16$ & $16$ \\
$8$ & $56$ & $56$  & $56$ \\
$16$ & $168$ & $182$ & $168$ \\
$32$ & $456$ & $506$  & $456$ \\
$64$ & $1160$ & $1298$  & $1160$ \\
$128$ & $2824$ & $3170$ &  $2824$ \\
$256$ & $6664$ & $7490$ & $6664$ \\
$512$ & $15368$ & $17282$ & $15368$ \\
$1024$ & $34824$ & $39170$ & $34824$ \\
$2048$ & $77832$ & $87554$ & $77832$ \\
\bottomrule
\end{tabular}
}
\end{table}



\section{Conclusions} \label{sec:finale}

The addition of an appropriate intermediate decomposition in the classical QFT algorithm produce a more efficient QFT algorithm, with the same
computational cost of celebrated split-radix 3add/3mul, but keeping the lower number of trigonometric costants of classical QFT versus split-radix 3add/3mul.
These characteristics make the improved QFT algorithm a good choise for $\CDFT$, $\RDFT$, $\DCT$, $\DST$ computation, in fixed point implementation, where a multiplication is slower than an addition, or in a parallel pipeline hardware implementation.
Moreover, an efficient `in place' implementation of improved QFT can be object of future research, since it is possible, but it is is not available yet.


\section{Acknoledgments}

I want to thank without implicating Michele Pasquini, Stefano Squartini and Francesco Piazza, who helped the author in revision and translation of this paper.


\begin{figure}[htb]
  \centering
  \includegraphics[width=1.3 \textwidth, angle=90]{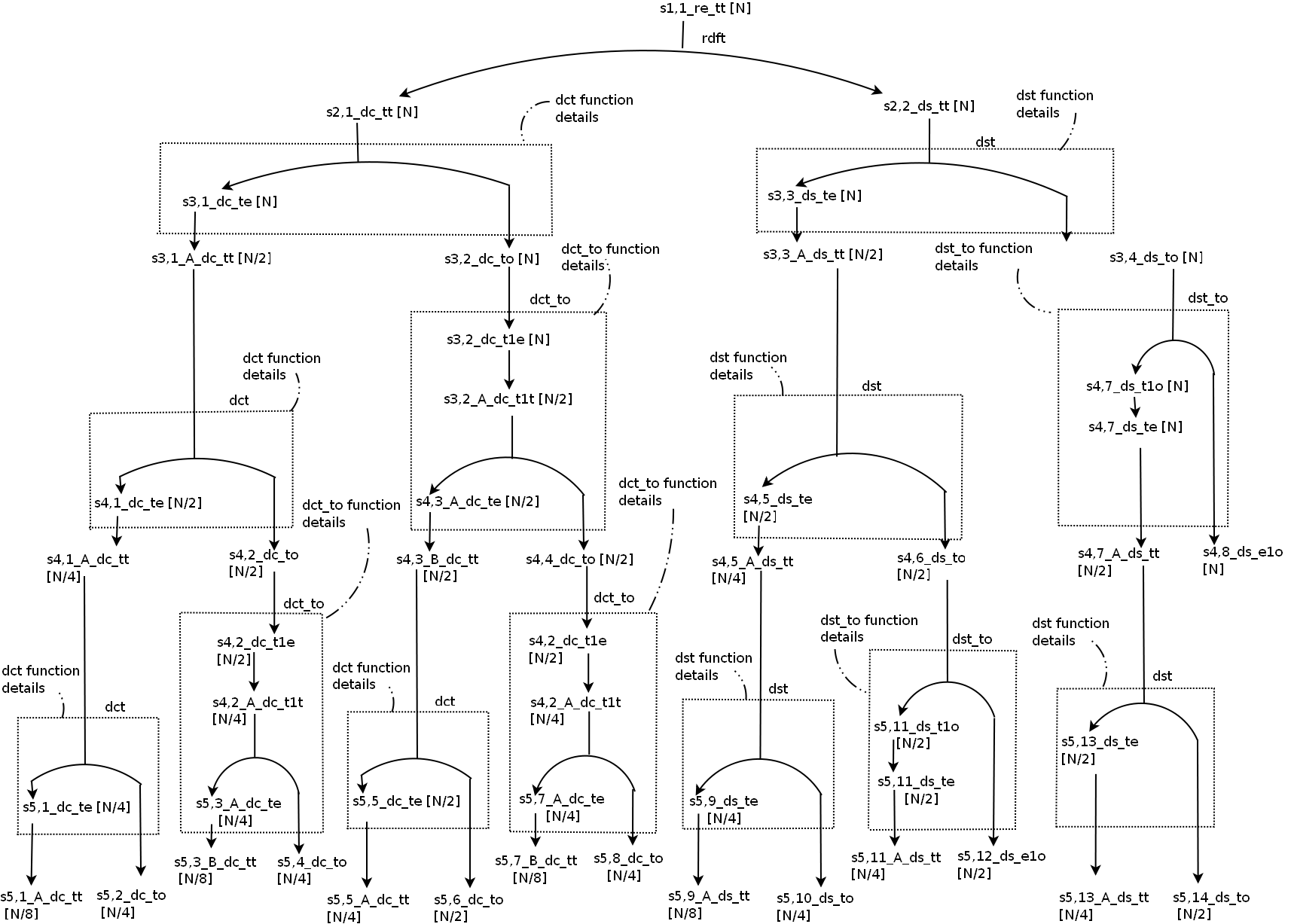}
  \caption{The decomposition tree of the classical QFT}
  \label{fig:classical_qft}
\end{figure}

\begin{figure}[htb]
  \centering
  \includegraphics[width=1.6 \textwidth, angle=90]{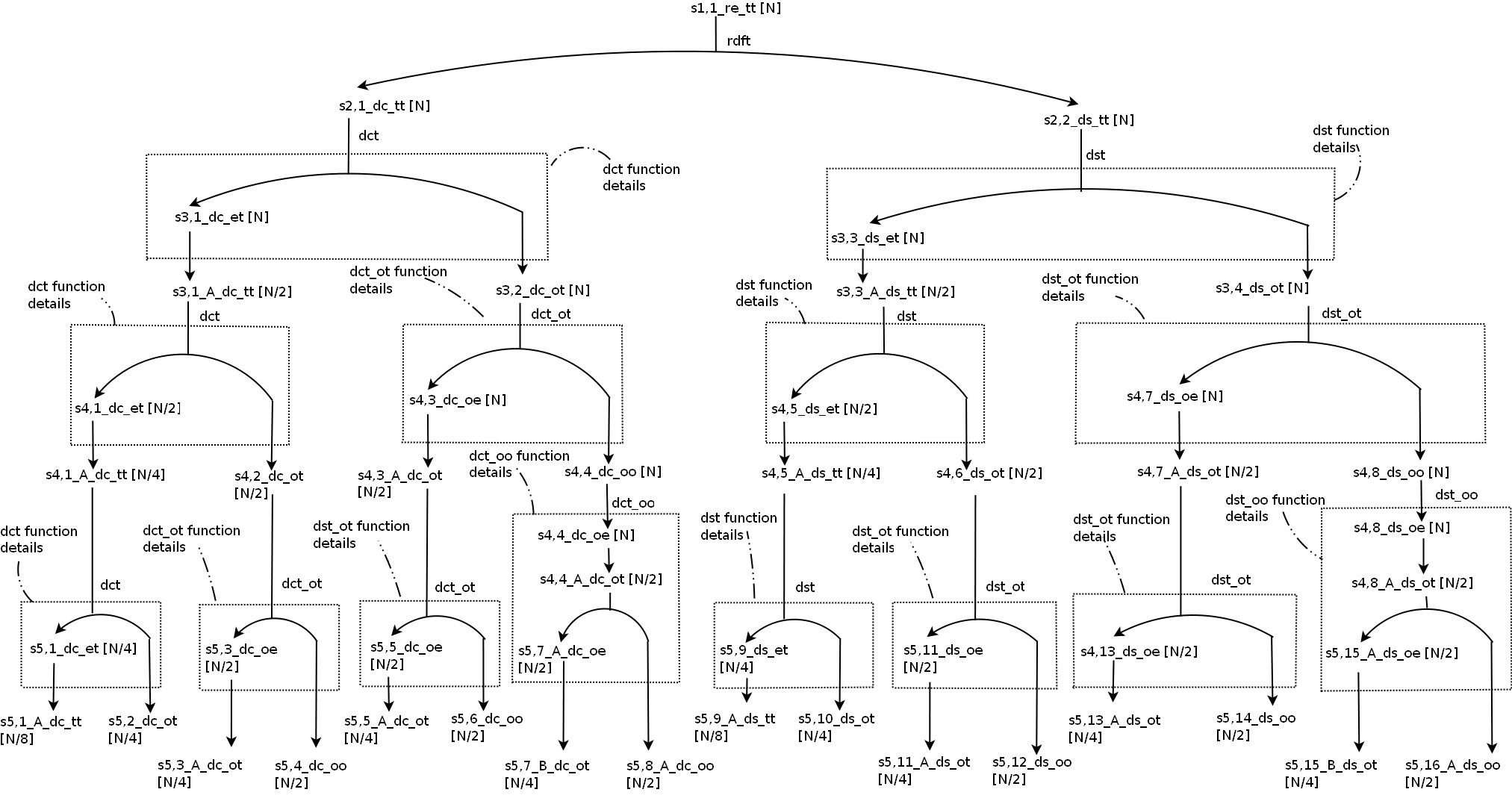}
  \caption{The decomposition tree of the improved QFT}
  \label{fig:improved_qft}
\end{figure}


\appendix


\begingroup
\refstepcounter{section}

\section*{Appendix: Advantages of signals types and notation} \label{sec:appendice}
 
As mentioned in sect. \ref{sec:notation}, the use of the `signal types' has numerous advantages.
In certain aspects these advantages are also reinforced by the particular notation used to indicate the various types of signals created.
These advantages are inherent three aspects: implementation of the theoretical algorithm in a programming language,  ideation and development of a recursive version of the algorithm, theoretical description of the algorithm.
Differently this approach has just a few disadvantages:
\begin{itemize}

\item
the reader have to spend time in comprehension of this atipical approach to describe FFT algorithms.

\item 
signal names are long (this disadvantage is specific to the particular notation used, but it is not intrinsic of signal types use)

\item
we need to keep an eye on Tab.\ref{tab:notation}, while we read the paper (also if we have used a nemonic notation).

\end{itemize}

\subsection{Advantages inherent implementation of the theoretical algorithm in a programming language}

Implementative aspects are usually neglected in the theoretical description of theoretical FFT algorithms, in literature.
The identification of signal types (referred to with the notation) already in the theoretical description of the algorithm, has the significant advantage to make the theoretical algorithm `ready to be implemented' for any reader who wants to implement the algorithm in the programming language he prefers.
This characteristic minimizes the time, and hence the cost, of implementation (coding) of the algorithm in any programming language that requires memory management to allocate the signals created by the algorithm into memory cells. 
In fact, all we need to manage the used signals (except mathematical details) is written in Tab.\ref{tab:notation}, \ref{tab:implementation}.
In general, this advantage is quantitatively more important if we implement many algorithms, or a few algorithms with many functions, or algorithms with many different signals but with a few signal types.
In particular, this advantage is realized in these aspects (during the writing phase of the code that implements the theoretical algorithm):
\begin{enumerate}

\item
we do not need to compute the number of array cells that serve to hold each stored signal, every time we handle a new signal, because it has already been made in  Tab.\ref{tab:notation}, provided in the theoretical description of algorithm.
This number of array cells is $max (ln (s), lk (s)))$.

\item
we just need to determine the matches between theoretical signal (temporal or frequency-domain) components and array of memory cells which contain residual temporal components or required frequency-domain components, inside the area of memory reserved for the signal, only for each used signal type, instead of as many times, as a signal type appears in the algorithm.
In fact the found matches are reusable each time an already previously used signal type appears.
Tab.\ref{tab:implementation} (which we obtain from Tab.\ref{tab:notation}) shows these matches for each signal type (these matches are valid only if we store indeces in growing order, for the programming languages, as Scilab-Malab, where the first cell of an array is $p=1$, where each cell of this array can contain a real or a complex value).

\item
the code is much more readable, since the array-signal terms let the reader of code to immediately identify all relevant characteristics of the signal stored in each array.

\item
the debugging time of code which implements the algorithm decreases, thanks to the use of Tab.\ref{tab:notation}, \ref{tab:implementation} .

\item
the time we spend to write comments on code decreases very much.
In fact:
\begin{itemize}

\item 
we can write comments in $dst\_oo$ ($dst\_ot$) function just copying them from the ones already written in $dct\_oo$ ($dct\_ot$) function, just substituting the $dc$ subscript with $ds$, thanks to the parallelism between these two functions, and thanks to the used notation for signal terms.

\item
we can write comments in $dst\_ot$ function just copying them from the ones already written in $dct\_cla$ function, just substituting the $sto\_n$ subscript, from `$t$' to `$o$', thanks to notation used for signal names, since these two functions apply the same chain of elaborations (only involved signal types change).

\end{itemize}

\end{enumerate}


\subsection{Advantages inherent the ideation and development of the recursive version of improved QFT}

The determination of the tipology of each created signal, has facilitated the development of the new improved QFT algorithm, in the theoretical stage.
In fact:
\begin{itemize}
\item
we immediatly know if a signal created in a function has the same characteristics of another signal already created before in another function (and therefore we can already know how to efficiently manage it, using the procedure already used in other functions).

\item
knowing the type of any created child signal lets us to control if every recursive function maintains the total output $ln$ ($lk$) parameters of the descendent signals (obtained summing the $ln$ ($lk$) parameters of the two signals) identical to the ones of the single input signal of the function.
This aspect is important, since the lack of this feature for $dct\_ot$ function makes the classical QFT nor efficient, nor in-place 
(as shown in sect. \ref{sec:memory_old_qft}).

\end{itemize}


\subsection{Advantages inherent the theoretical description of the algorithm in a paper}

This category includes these benefits:
\begin{enumerate}

\item
high intelligibility of the description of algorithms, since:
\begin{itemize}

\item
each signal name (sequence of symbols) has the same meaning everywhere in this paper.

\item
the reader knows any detail about the way the algorithm acts, in every step of the algorithm. 
In fact, in any mathematical relationship described, the reader can obtain any relevant information about signals involved (for example in eq.(\ref{equ:qft_26})), from their names, using Tab.\ref{tab:notation}.  

\item
used notation make the characteristics of any signal type, mnemonic.

\item
signal types explicit the subtleties (the differences in the $sto\_n$ and $sto\_k$ groupings of signals created by the two algorithms shown in this paper),  that cause the greater efficiency (both as a computational cost, and as a memory requirement) of the improved QFT, compared to the classical version of the QFT (details are explained in sect. \ref{sec:memory_new_qft}).

\end{itemize}

\item
Compactness in exposition of all details of any used elaboration.
In fact:
\begin{itemize}

\item
The use of signal types lets to detect identical functions in distinct algorithms, thus to avoid to describe them several times in the paper. 
Understanding if two functions are identical is not trivial.
In fact using the same chain of elaborations is not enough to make two functions identical: they have to apply to the same signal type too.
For example, $dct\_cla$ and $dct\_ot$ functions described in this paper use the same chain of elaborations, but they are different functions, since they apply to different input signals ($s_{dc\_tt}$ and $s_{ds\_ot}$ respectively).

\item
Understanding the used mathematical relationships (for example eq. (\ref{equ:qft_26})) does not require additional comments written in natural language, if any signal is described by the notation 
(except for indication of the periodization of any signal present). 
In fact, each additional line of text concerning the description of the signals involved is superfluous. 
For example, we don't need to comment: `for this signal we are interested  in calculating frequency-domain components only for even harmonics $ k \in \{0,2,\dots,(\frac{N}{4})\}$, through $\DCT$', because we can  directly deduce it from Tab.\ref{tab:notation}.

\item
the particular notation used to describe signal types lets us to describe a function just stating it is analogous to another function, and describing how signal types change, without loosing any detail. 
For example, in sect. \ref{sec:new_qft_dst_functions}, we can describe the $dst\_ot$ and $dst\_oo$ functions, using only 4 lines of text, stating these functions use the same chain of elaborations used in $dct\_ot$ and $dct\_oo$ functions respectively, but applied to different root signal types, and thus we just need to replace the symbol $dc$ with the symbol $ds$ in any created signal type.
It is important to note that the statement `we just need to replace everywhere the symbol `$dc$' with symbol `$ds$' ' gives us many more informations than the sentence `just replace the $\DCT$ calculation with the $\DST$ calculation'.
In fact the last sentence doesn't inform us if this analogia keeps inalterated, or changes, the $sto\_n$ or $sto\_k$ groups, of signals involved in the two functions. 

\end{itemize}

\item
The possibility of using a graphical and intuitive description of the algorithm: the tree decomposition (see Fig.\ref{fig:classical_qft} and \ref{fig:improved_qft}), where we report the concatenation of the processing performed, and the signal types to which they apply.
Let us consider that the informations conveyed from the decomposition tree would be much lower if we insert signals with random names in this graph.

The decomposition tree is useful because it gives us the view of how to manage the memory area reserved for data (dividing it among the various signals, in each level of decomposition), if combined with Tab.\ref{tab:notation}. 
In fact, if we want, we can sort the signals in the memory area, in the same order they occur in the decomposition tree. Thus we know where each signal is located in memory.

\end{enumerate}


\begin{table}[tb]
\caption{a possible matching between (theoretical type of signal) and (array of memory cells), in an implementation where each signal is stored into a contiguous sequence of cell (array), where indeces $n$ as $k$ are stored in growing order, and the the first cell of an array has index $p=1$}
\label{tab:implementation}
\centering
\scalebox{0.9}
{
\begin{tabular}{ccc}
\toprule
signal type & temporal signal-array matching & transformed signal-array matching \\
\midrule
$s_{cx\_tt}$ & $s_{cx\_tt}(n)=s_{cx\_tt\_arr}(n+1)$ & $S_{cx\_tt}(k)=S_{cx\_tt\_arr}(k+1)$ \\
$s_{re\_tt}$ & $s_{re\_tt}(n)=s_{re\_tt\_arr}(n+1)$ & $S_{re\_tt}(k)=S_{re\_tt\_arr}(k+1)$ \\
$s_{dc\_tt}$ & $s_{dc\_tt}(n)=s_{dc\_tt\_arr}(n+1)$ & $S_{dc\_tt}(k)=S_{dc\_tt\_arr}(k+1)$ \\
$s_{dc\_et}$ & $s_{dc\_et}(n)=s_{dc\_et\_arr}(\frac{n+2}{2})$ & $S_{dc\_et}(k)=S_{dc\_et\_arr}(k+1)$ \\
$s_{dc\_ot}$ & $s_{dc\_ot}(n)=s_{dc\_ot\_arr}(\frac{n+1}{2})$ & $S_{dc\_ot}(k)=S_{dc\_ot\_arr}(k+1)$ \\
$s_{dc\_te}$ & $s_{dc\_te}(n)=s_{dc\_te\_arr}(n+1)$ & $S_{dc\_te}(k)=S_{dc\_te\_arr}(\frac{k+2}{2})$ \\
$s_{dc\_to}$ & $s_{dc\_to}(n)=s_{dc\_to\_arr}(n+1)$ & $S_{dc\_to}(k)=S_{dc\_to\_arr}(\frac{k+1}{2})$ \\
$s_{dc\_oe}$ & $s_{dc\_oe}(n)=s_{dc\_oe\_arr}(\frac{n+1}{2})$ & $S_{dc\_oe}(k)=S_{dc\_oe\_arr}(\frac{k+2}{2})$ \\
$s_{dc\_oo}$ & $s_{dc\_oo}(n)=s_{dc\_oo\_arr}(\frac{n+1}{2})$ & $S_{dc\_oo}(k)=S_{dc\_oo\_arr}(\frac{k+1}{2})$ \\
$s_{dc\_t_{1} e}$ & $s_{dc\_t_{1} e}(n)=s_{dc\_t_{1} e\_arr}(n+1)$ & $S_{dc\_t_{1} e}(k)=S_{dc\_t_{1} e\_arr}(\frac{k+2}{2})$ \\
$s_{dc\_t_{1} t}$ & $s_{dc\_t_{1} t}(n)=s_{dc\_t_{1} t\_arr}(n+1)$ & $S_{dc\_t_{1} t}(k)=S_{dc\_t_{1} t\_arr}(k+1)$ \\
$s_{ds\_tt}$ & $s_{ds\_tt}(n)=s_{ds\_tt\_arr}(n)$ & $S_{ds\_tt}(k)=S_{ds\_tt\_arr}(k)$ \\
$s_{ds\_et}$ & $s_{ds\_et}(n)=s_{ds\_et\_arr}(\frac{n}{2})$ & $S_{ds\_et}(k)=S_{ds\_et\_arr}(k)$ \\
$s_{ds\_ot}$ & $s_{ds\_ot}(n)=s_{ds\_ot\_arr}(\frac{n+1}{2})$ & $S_{ds\_ot}(k)=S_{ds\_ot\_arr}(k)$ \\
$s_{ds\_te}$ & $s_{ds\_te}(n)=s_{ds\_te\_arr}(n)$ & $S_{ds\_te}(k)=S_{ds\_te\_arr}(\frac{k}{2})$ \\
$s_{ds\_to}$ & $s_{ds\_to}(n)=s_{ds\_to\_arr}(n)$ & $S_{ds\_to}(k)=S_{ds\_to\_arr}(\frac{k+1}{2})$ \\
$s_{ds\_oe}$ & $s_{ds\_oe}(n)=s_{ds\_oe\_arr}(\frac{n+1}{2})$ & $S_{ds\_oe}(k)=S_{ds\_oe\_arr}(\frac{k}{2})$ \\
$s_{ds\_oo}$ & $s_{ds\_oo}(n)=s_{ds\_oo\_arr}(\frac{n+1}{2})$ & $S_{ds\_oo}(k)=S_{ds\_oo\_arr}(\frac{k+1}{2})$ \\
$s_{ds\_t_{1} o}$ & $s_{ds\_t_{1} o}(n)=s_{ds\_t_{1} o\_arr}(n)$ & $S_{ds\_t_{1} o}(k)=S_{ds\_t_{1} o\_arr}(\frac{k+1}{2})$ \\
$s_{ds\_e_{1} o}$ & $s_{ds\_e_{1} o}(n=\frac{N}{4})=s_{ds\_e_{1} o\_arr}(1)$ & $S_{ds\_e_{1} o}(k=1)=S_{ds\_e_{1} o\_arr}(1)$ \\
\bottomrule
\end{tabular}
}
\end{table}

\endgroup


\bibliographystyle{plain}

\bibliography{biblio}

\end{document}